\documentstyle[editedvolume]{crckapb}

\def\aj#1{{\it Astron. J.,} {\bf #1}}
\def\apj#1{{\it Astrophys. J.,} {\bf #1}}
\def\apjss#1{{\it Astrophys. J. Suppl. Ser.,} {\bf #1}}
\def\aa#1{{\it Astron. \& Astrophys.,} {\bf #1}}
\def\araa#1{{\it Annu. Rev. Astron. \& Astrophys.,} {\bf #1}}

\def\aas#1{{\it Astron. \& Astrophys. Suppl.,} {\bf #1}}
\def\mn#1{{\it Mon. Not. Roy. Astr. Soc.,} {\bf #1}}
\def\kfnt#1#2{{\it Kinematika i Fizika Nebesnykh Tel,} {\bf #1}, {No. #2}}
\def\ass#1{{\it Astrophys. \& Sp. Sci.,} {\bf #1}}

\def\pasp#1{{\it Publ. of Astr. Soc. of Pac.,} {\bf #1}}

\def\be{\begin{equation}}
\def\ee{\end{equation}}

\def\bea{\begin{eqnarray}}
\def\eea{\end{eqnarray}}

\def\eg{{\it e.g.}}
\def\ie{{\it i.e.}}
\def\etal{{\it et al.}}

\def\vsp{\vspace{8.0cm}}

\def\insertplot#1{
\includegraphics{#1} }

\begin{opening}

\title{Modeling the star formation in galaxies \protect \\
using the Chemo - dynamical SPH code}

\subtitle{Chemo - dynamical SPH code}

\author{Peter BERCZIK}

\institute{Main Astronomical Observatory of \\
           Ukrainian National Academy of Sciences \\
           252650, Golosiiv, Kiev-022, Ukraine \\
           e-mail: {\tt berczik@mao.kiev.ua}}

\end{opening}

\runningtitle{Chemo - dynamical SPH code}

\begin{document}






\begin{abstract}

     A new Chemo - Dynamical Smoothed Particle Hydrodynamic
(CD - SPH) code is presented. The disk galaxy is described
as a multi - fragmented gas and star system, embedded in a
cold dark matter halo. The star formation (SF) process,
SNII, SNIa and PN events as well as chemical enrichment of
gas have been considered within the framework of the
standard SPH model. Using this model we describe the
dynamical and chemical evolution of triaxial disk - like
galaxies. It is found that such approach provides a
realistic description of the process of formation, chemical
and dynamical evolution of disk galaxies over a
cosmological timescale.

\end{abstract}



\section{Introduction}

     The dynamical and chemical evolution of galaxies is one of
the most interesting and complex problems. Naturally,
galaxy formation is tightly connected with the process of
large - scale structure formation in the Universe. The main
role in the scenario of large - scale structure formation
seems to be played by the dark matter. It is believed that
the Universe was seeded at some early epoch with small
density fluctuations of dark non - baryonic matter, and the
evolving distribution of these dark halos provides the
arena for galaxy formation. Galaxy formation itself
involves collapse of baryons within potential wells of dark
halos \cite{WR78}. The properties of forming galaxies
depend on the amount of baryonic matter, that can be
accumulated in such halos, and the efficiency of star
formation. The observational support for this scenario of
galaxy formation comes from the recent COBE detection of
fluctuations in the microwave background \eg~
\cite{BBHMSW93}.

     The investigation of the process of galaxy formation is a highly
complex subject requiring many different approaches. The
formation of self - gravitating inhomogeneities of
protogalactic size, the ratio of baryonic and non -
baryonic matter \cite{BBKSz86,WS79,P93,D95} the origin of
the protogalaxy's initial angular momentum
\cite{VH89,ZQS88,EL95,SB95}, the protogalaxy's collapse and
its subsequent evolution are usually considered as
separated problems. The recent advances in computer
technology and numerical methods have made possible the
detailed modeling of baryon matter dynamics in the universe
dominated by collisionless dark matter and, therefore, the
detailed gravitational and hydrodynamical description of
the formation and evolution of galaxies. The most
sophisticated models additionally include radiative
processes, star formation and supernova feedback \eg~
\cite{K92,SM94,FB95}.

     However, results of numerical simulations are
essentially affected by the star formation algorithm
incorporated into modeling techniques. The star formation
and accompanied processes are still not well understood on
either small or large spatial scales. The star formation
algorithm by which the gas material is converted into stars
can be based only on simple theoretical assumptions or on
results of empirical observations of nearby galaxies. The
other most important effect of star formation on the global
evolution of a galaxy is caused by a large amount of energy
released in supernova explosions and stellar winds.

     Among numerous methods developed for the modeling of
complex three dimensional hydrodynamic phenomena Smoothed
Particle Hydrodynamics (SPH) is one of the most popular
\cite{M92}. Its Lagrangian nature allows to be combined
easily with fast N - body algorithms, making it very
suitable for simultaneous description of complex dynamics
of gas - stellar system \cite{FB95}. As an example of such
combination, the TREE - SPH code \cite{HK89,NW93} can be
mentioned, which was successfully applied to the detailed
modelling of disk galaxy mergers \cite{MH96} and galaxy
formation and evolution \cite{K92}. The second good example
is a GRAPE - SPH code \cite{SM94,SM95} which was
successfully used to model of the evolution of disk galaxy
structure and kinematics.

     In recent few years, we had an high number of excellent papers
concerning the complex SPH modeling of galaxy formation and
evolution \cite{RVN96,CLC97}. In our code, we propose new
"energetic" criteria for SF, and make a more realistic
account of returned chemically enriched gas fraction via
SNII, SNIa and PN events.

     The simplicity and numerical efficiency of the SPH method were the
main reasons why we chose this technique for the modeling
of the evolution of complex, triaxial protogalactic
systems. We used our own modification of the hybrid N -
body/SPH method \cite{BerK96,Ber98}, which we call the
chemo - dynamical SPH (CD - SPH) code.

     The "stars" are included into the standard SPH
algorithm as the N - body collisionless system of particles, which
can interact with the gas component only through the gravitation
\cite{K92}. The star formation process and supernova explosions
are included into the scheme in the manner proposed by
Raiteri \etal~ (1996), but with our own modifications.


\section{The CD - SPH code}

\subsection{The SPH code}

     Continuous hydrodynamic fields in SPH are described by the
interpolation functions constructed from the known values of these
functions at randomly positioned particles. In this case the mean
value of a physical field $ f( {\bf r} ) $ at the point $ {\bf r}
$ can be written as \cite{M92}:

\be
   < f( {\bf r} ) > = \int f( {\bf r'} ) \cdot
                           W( {\bf r} - {\bf r'}; h ) d{\bf r'},
   \label{eq:def_f}
\ee

where $ W( {\bf r} - {\bf r'}; h ) $ is the kernel function and $
h $ is the softening constant.

     The function $ W({\bf r} - {\bf r'};h) $ is strongly peaked at $
\mid {\bf r} - {\bf r'} \mid  = 0 $, and we can consider it
without any loss of generality as belonging to the class of even
functions. In this case it is not difficult to demonstrate
\cite{HK89} that the average value $ < f( {\bf r} ) > $ represents
the real function $ f( {\bf r} ) $, with an error not higher than
$ O(h^2) $. Consider a fluid with the density $ \rho({\bf r}) $,
then we can rewrite (\ref{eq:def_f}) in the form:

\be
   < f( {\bf r} ) > = \int \frac{f({\bf r'}) }{\rho({\bf r'})} \cdot
                           W( {\bf r} - {\bf r'}; h ) \cdot
                           \rho({\bf r'}) d{\bf r'} .
   \label{eq:def_fff}
\ee

     Let us imagine that $ f({\bf r}) $ and $ \rho({\bf r}) $ are known
only at $ N $ discrete points $ {\bf r}_i $. Then, equation
(\ref{eq:def_fff}) gives:

\be
    f_i = \sum_{j=1}^{N} m_j \cdot \frac{f_j}{\rho_j} \cdot W_{ij}.
   \label{eq:def_f_fi}
\ee

     Here $ f_i \equiv f( {\bf r}_i ) $, $ \rho_i \equiv \rho({\bf r}_i)
$, $ W_{ij} \equiv W( {\bf r}_i - {\bf r}_j; h ) $ and $ m_i $ is
the mass of particle $ i $. Using (\ref{eq:def_f_fi}), we
approximate the hydrodynamic field by an analytical function which
is differentiable as many times as the kernel $ W_{ij} $.
Following
Monaghan \& Lattanzio (1985) we use for the kernel function $
W_{ij} $ the spline expression in the form:

\bea
   W_{ij} = \frac{1}{\pi h^3}
            \left\{
            \begin{array}{lllll}
            1 - \frac{3}{2} u_{ij}^2 + \frac{3}{4} u_{ij}^3, & \mbox{ if $0 \leq u_{ij} < 1$}, \\
                                                                                                            \\
            \frac{1}{4} (2  - u_{ij})^3,                     & \mbox{ if $1 \leq u_{ij} < 2$}, \\
                                                                                                            \\
            0,                                               & \mbox{ otherwise }. \\
            \end{array}
            \right.
   \label{eq:def_w}
\eea

     Here $ u_{ij} = r_{ij}/h $.

     The local resolution in SPH is defined by the chosen smoothing
length $ h $. If it is the same for all points, the hydrodynamic
field will be evidently approximated more smoothly in the regions
where the points lie more closely, \ie~ where the density is
higher.

     To achieve the same level of accuracy for all points in the
fluid it is necessary to use a spatially variable smoothing
length. In this case each particle has its individual value
of $ h $. Following
Hernquist \& Katz (1989), we write instead of (\ref{eq:def_f}):

\be
   < f( {\bf r} ) > = \int f( {\bf r'} ) \cdot \frac{1}{2} \cdot
   [W( \delta{\bf r}; h) + W( \delta{\bf r}; h')] d{\bf r'}.
   \label{eq:def_f_new}
\ee

     Here $ \delta{\bf r} \equiv {\bf r} - {\bf r'} $ , $ h \equiv
h({\bf r}) $ and $ h' \equiv h({\bf r'}) $. For the density $
\rho({\bf r}) $ it gives

\be
   < \rho({\bf r}_i) > = \sum_{j=1}^{N} m_j \cdot \frac{1}{2} \cdot
   [W(r_{ij}; h_i) + W(r_{ij}; h_j)].
   \label{eq:def_rho_new}
\ee

     Other hydrodynamic functions are written in the same manner.

     In our calculations the values of $ h_i $ were determined from the
condition that the number of particles $ N_B $ in the
neighborhood of each particle within the $ 2 \cdot h_i $
remains constant \cite{MH96}. The value of $ N_B $ is
chosen such that a certain fraction from the total number
of "gas" particles $ N $ affects the local flow
characteristics \cite{HV91}. If the defined $ h_i $ become
smaller than the minimal smoothing length $ h_{min} $, we
set the value $ h_i = h_{min} $. For "star" particles (with
Plummer density profiles) we use, accordingly, the fixed
gravitational smoothing lengths $ h_{star} $.

     If  the density is computed according to equation
(\ref{eq:def_rho_new}), then the continuity equation is satisfied
automatically. Equations of motion for particle $ i $ are

\be
   \frac{d{\bf r}_i}{dt} = {\bf v}_i,
   \label{eq:def_r}
\ee

\be
   \frac{d{\bf v}_i}{dt} = -\frac{\nabla_i P_i}{\rho_i} +
                    {\bf a}^{vis}_{i} -
                    \nabla_i \Phi_i -
                    \nabla_i \Phi^{ext}_i,
   \label{eq:def_v}
\ee

     where $ P_i $ is the pressure, $ \Phi_i $ is the self gravitational
potential, $ \Phi^{ext}_i $ is a gravitational potential of
possible external halo, $ {\bf a}^{vis}_{i} $ is an artificial
viscosity term. Using equations (\ref{eq:def_f_new}) and
(\ref{eq:def_rho_new}) we can write:

\be
   \frac{\nabla_i P_i}{\rho_i} = \sum_{j=1}^{N} \frac{m_j}{2}
           (\frac{P_i}{\rho_i^2}+\frac{P_j}{\rho_j^2})
           \nabla_i [W(r_{ij};h_i) + W(r_{ij};h_j)].
   \label{eq:def_grad_p}
\ee

     The artificial viscosity term $ {\bf a}^{vis}_{i} $ is introduced
so as to describe the flows with shock waves. In present
calculations the viscous acceleration is introduced by replacing $
(P_i/\rho_i^2 + P_j/\rho_j^2) $ in equation (\ref{eq:def_grad_p})
by $ (P_i/\rho_i^2 + P_j/\rho_j^2) \cdot (1 + \pi_{ij}) $. The
expression for $ \pi_{ij} $ has the form \cite{HVC91}:

\be
   \pi_{ij} = -\alpha \cdot \mu_{ij} + \beta \cdot \mu_{ij}^2,
   \label{eq:def_pi}
\ee

     where $ \alpha $ and $ \beta $ are constants, and $ \mu_{ij} $ is
defined by the relation:

\bea
   \mu_{ij} =
            \left\{
            \begin{array}{lll}
            h_{ij}/c_{ij} \cdot ({\bf v}_{ij} {\bf r}_{ij}) /
            (r^2_{ij} + n^2 h_{ij}^2),  & \mbox{if $ ({\bf v}_{ij} {\bf r}_{ij}) < 0 $}, \\
                                                                          \\
            0,                                                 & \mbox{otherwise}. \\
            \end{array}
            \right.
   \label{eq:def_mu}
\eea

     Here $ c_{ij} = (c_i + c_j)/2 $ is the sound speed averaged for
points $ i $ and $ j $, $ h_{ij} = (h_i + h_j)/2 $, $ {\bf r}_{ij}
= ({\bf r}_i - {\bf r}_j) $ and $ {\bf v}_{ij} = ({\bf v}_i - {\bf
v}_j) $ is the relative velocity vector for the points $ i $ and $
j $. The term $ n^2 \cdot h_{ij}^2 $ is inserted to prevent
divergences when $ r_{ij} = 0 $ and it should be small enough. The
constant $ n $ was set equal to $ 0.1 $. For the constant $ \alpha
$ and $ \beta $ in (\ref{eq:def_pi}) the values $ \alpha = 1 $ and
$ \beta = 2 $ give good results in a wide range of the Mach
numbers \cite{M92}.

     By using equation (\ref{eq:def_w}), one gets for the gravitational
acceleration \cite{HV91}:

\be
   - \nabla_i \Phi_i = - \frac{1}{2} \cdot \sum_{j=1}^{N} G \cdot m_j \cdot
                         \frac{{\bf r}_{ij}}{r_{ij}^3} \cdot
                         [g_i + g_j],
   \label{eq:def_Phi_1}
\ee

    where:

\bea
   g_k = \left\{
            \begin{array}{lllll}
            \frac{4}{3} u_k^3 - \frac{6}{5} u_k^5 + \frac{1}{2} u_k^6,         & \mbox{$0 \leq u_k < 1$}, \\
                                                                                                                                  \\
            -\frac{1}{15} + \frac{8}{3} u_k^3 - 3 u_k^4 + \frac{6}{5} u_k^5 - \frac{1}{6} u_k^6, & \mbox{$1 \leq u_k < 2$}, \\
                                                                                                                                  \\
            1,                                                                                   & \mbox{otherwise}, \\
            \end{array}
            \right.
\eea

     and $ u_k = r_{ij}/h_k $.

     When isothermal flows are considered, the system of equations is
closed by adding the equation of state:

\be
   P_i = \rho_i \cdot c_i^2,
   \label{eq:def_P}
\ee

     where $ c_i $ is the isothermal speed of sound.

     In the particle representation for complex hydrodynamic flows the
energy equation has the form:

\bea
   \frac{du_i}{dt} & = & \sum_{j=1}^{N} \frac{m_j}{4}
           (\frac{P_i}{\rho_i^2}+\frac{P_j}{\rho_j^2})
           (1 + \pi_{ij}) \times \nonumber \\
    & &    \times \nabla_i [W(r_{ij};h_i) + W(r_{ij};h_j)]
           {\bf v}_{ij} + \frac{\Gamma_i - \Lambda_i}{\rho_i}.
   \label{eq:def_du}
\eea

     Here $ u_i $ is the specific internal energy of particle $ i $.
The term  $ (\Gamma_i - \Lambda_i)/\rho_i $ accounts for
nonadiabatic processes not associated with the artificial
viscosity (in our calculations $ \Gamma_i \equiv 0 $). We present
the radiative cooling in the form:

\be
   \Lambda_i = \Lambda_i(u_i, \rho_i) = \Lambda^{*}_i(T_i) \cdot n^2_i,
   \label{eq:lam_i}
\ee

     where $ n_i $ is the hydrogen number density and $ T_i $ is the
temperature. To follow its subsequent thermal behaviour in
numerical simulations we use an analytical approximation of the
standard cooling function $ \Lambda^{*}(T) $ for an optically thin
primordial plasma in ionization equilibrium \cite{DM72,KG91}. Its
absolute cutoff temperature is set equal to $ 10^4 $ K.

     If $ \log T \le 6.2 $:

\bea
   \Lambda^{*}(T) & = &   10^{-28} \cdot [10^{-0.1 - 1.88 (5.23 - \log T)^4} + \nonumber \\
                  &   & + 10^{-1.7-0.2 (6.2 - \log T)^4}] \; {\rm J/s},
   \label{eq:lam_T1}
\eea

     else

\bea
   \Lambda^{*}(T) & = &   10^{-28} \cdot [10^{-0.1 - 1.88 (5.23 - \log T)^4} + \nonumber \\
                  &   & + 10^{-1.7}] \; {\rm J/s},
   \label{eq:lam_T2}
\eea

     The equation of state must be added to close the system.

\be
   P_i = \rho_i \cdot (\gamma - 1) \cdot u_i,
   \label{eq:def_P_i_1}
\ee

     where $ \gamma = 5/3 $ is the adiabatic index.


\subsection{Time integration}

     To solve the system of equations (\ref{eq:def_r}), (\ref{eq:def_v})
and (\ref{eq:def_du}) we use the leapfrog integrator \cite{HK89}.
This system of equations has the form:

\bea
   \left\{
   \begin{array}{lllll}
   d{\bf r}_i/dt = {\bf v}_i,                            \\
                                                         \\
   d{\bf v}_i/dt = {\bf a}_i(h, \rho, T, c, P, {\bf r}, {\bf v}), \\
                                                         \\
   du_i/dt = \dot u_i(h, \rho, T, c, P, {\bf r}, {\bf v}).        \\
   \end{array}
   \right.
   \label{eq:def_dr_dv_du}
\eea

     The time step $ \delta t_i $ for each particle depends on the
particle's acceleration $ {\bf a}_i $ and velocity $ {\bf v}_i $,
as well as on viscous forces. To define $ \delta t_i $ we use the
relation \cite{HV91}:

\be
   \delta t_i = C_n \min [\sqrt{\frac{h_i}{\mid {\bf a}_i \mid}},
                \frac{h_i}{\mid {\bf v}_i \mid},
                \frac{h_i}{s_i}],
   \label{eq:def_dt}
\ee

     where $ s_i = c_i \cdot (1 + \alpha + 0.68 \cdot \beta \cdot
\max_{j} \mid \mu_{ij} \mid) $ and $ C_n $ is the Courant's
number. We adopt $ C_n = 0.1 $. The minimum time step value is set
equal to:

\be
\Delta t = \min_{i} \; (\delta t_i), \ee

     for all particles. The integrator has a second order accuracy in
the time step $ \Delta t $. As the first step for particle $ i $
we define:

\bea
  \left\{
  \begin{array}{lllll}
  {\bf r}^{n+\frac{1}{2}}_i = {\bf r}^{n}_i + 0.5 \cdot \Delta t \cdot {\bf v}^{n}_i, \\
  \\
  {\bf v}^{n+\frac{1}{2}}_i = {\bf v}^{n}_i + 0.5 \cdot \Delta t \cdot {\bf a}^{n}_i, \\
  \\
  u^{n+\frac{1}{2}}_i       = u^{n}_i       + 0.5 \cdot \Delta t \cdot \dot u^{n}_i. \\
  \end{array}
  \right.
  \label{eq:def_n05}
\eea

     After that:

\bea
  \left\{
  \begin{array}{lllll}
  {\bf r}^{n+1}_i = {\bf r}^{n}_i + \Delta t \cdot {\bf v}^{n+\frac{1}{2}}_i + O(\Delta t^3), \\
  \\
  {\bf v}^{n+1}_i = {\bf v}^{n}_i + \Delta t \cdot {\bf a}^{n+\frac{1}{2}}_i + O(\Delta t^3), \\
  \\
  u^{n+1}_i       = u^{n}_i + \Delta t \cdot \dot u^{n+\frac{1}{2}}_i        + O(\Delta t^3). \\
  \end{array}
  \right.
  \label{eq:def_n1}
\eea

     We have  carried  out \cite{BerK93} a large series of test
calculations to check that the code is correct, the conservation
laws are obeyed and the hydrodynamic fields are represented
adequately. These tests have shown very good results.


\subsection{The star formation algorithm}

     It is well known that star formation (SF) regions
are tightly connected to giant molecular complexes,
especially to regions, which reach some threshold for
dynamical instabilities. The overall picture of star
formation seems to be understood, but the detailed physics
of star formation and accompanying processes on either
small or large scales remains sketchy \cite{L69,S87,LRD92}.

     For describing of the process of conversion of gas
material into stars we modify the standard SPH star
formation algorithm \cite{K92,NW93}, taking into account
the presence of chaotic motions in the gaseous environment
and the time lag between the initial development of
suitable conditions for star formation and star formation
itself \cite{BerK96,Ber98}. The first reasonable
requirement incorporated into this algorithm allows
selecting "gas" particles that are potentially eligible to
form stars. It states that in the separate "gas" particle
the SF can start if the absolute value of the "gas"
particle's gravitational energy exceeds the sum of its
thermal energy and energy of chaotic motions:

\be
   \mid E_i^{gr} \mid > E_i^{th} + E_i^{ch}.
  \label{eq:crit}
\ee

     Gravitational and thermal energies and energy of chaotic motions
for the "gas" particle $ i $ in model simulation are defined as:

\bea
  \left\{
  \begin{array}{lllll}
  E_i^{gr} = - \frac{3}{5} \cdot G \cdot m_i^2/h_i,    \\
                                                       \\
  E_i^{th} = \frac{3}{2} \cdot m_i \cdot c_i^2,        \\
                                                       \\
  E_i^{ch} = \frac{1}{2} \cdot m_i \cdot \Delta v_i^2. \\
  \end{array}
  \right.
  \label{eq:energy}
\eea

     Where $ c_i = \sqrt{\Re \cdot T_i / \mu} $ is the isothermal sound
speed of particle $ i $. We set $ \mu = 1.3 $ and define the
chaotic or "turbulent" square velocities near particle $ i $ as:

\be
  \Delta v_i^2 = \sum_{j=1}^{N_{B}} m_j \cdot ({\bf v}_j - {\bf v}_c)^2 /
                 \sum_{j=1}^{N_{B}} m_j,
\ee

    where:

\be
  {\bf v}_c = \sum_{j=1}^{N_{B}} m_j \cdot {\bf v}_j /
              \sum_{j=1}^{N_{B}} m_j.
\ee

     For practical reasons, it is useful to define a critical
temperature for SF onset in particle $ i $ as:

\be
  T^{crit}_i = \frac{\mu}{3 \Re} \cdot
               ( \frac{8}{5} \cdot \pi \cdot G \cdot \rho_i \cdot h_i^2
               - \Delta v_i^2 ).
  \label{eq:t_crit}
\ee

     Then if the temperature of the "gas" particle $ i $,
drops below the critical one, SF can proceed.

\be
  T_i < T^{crit}_i.
  \label{eq:crit_T}
\ee

    We think that requirement (\ref{eq:crit}), or in an other form
(\ref{eq:crit_T}), is only a necessary one. It seems reasonable
that the chosen "gas" particle will produce stars only if the
above condition will hold over the time interval exceeding its
free - fall time $ t_{ff} = \sqrt {3 \cdot \pi / ( 32 \cdot G
\cdot \rho ) } $. This condition is based on the well known fact
that due to gravitational instability all substructures of
collapsing system are formed on such timescale. Using it we
exclude from the consideration transient structures, that are
destroyed by the tidal action of surrounding matter. This last
condition we assume to be a sufficient one.

     We also define which "gas" particles remain cool, \ie~ $ t_{cool} <
t_{ff} $. We rewrite this condition in the manner presented in
Navarro \& White (1993): $ \rho_i > \rho_{crit} $. Here we use the
value of $ \rho_{crit} = 0.03 $ cm$^{-3}$.

     When the collapsing particle $ i $ is defined, we create the new
"star" particle with mass $ m^{star} $ and update the "gas"
particle $ m_i $ using these simple equations:

\bea
  \left\{
  \begin{array}{lll}
  m^{star} = \epsilon \cdot m_i,  \\
                                  \\
  m_i = (1 - \epsilon) \cdot m_i. \\
  \end{array}
  \right.
  \label{eq:m}
\eea

     Here $ \epsilon $, defined as the global efficiency of star
formation, is the fraction of gas converted into stars according
to the appropriate initial mass function (IMF). The typical values
for SF efficiency in our Galaxy on the scale of giant molecular
clouds are in the range $ \epsilon \approx 0.01 - 0.4 $
\cite{DIL82,WL83}. But it is still a poorly known quantity. In
numerical simulation it is the model parameter which has to be
checked by comparison of numerical simulation results with
available observational data. Here we define $ \epsilon $ as:

\be
  \epsilon = 1 - (E_i^{th} + E_i^{ch})/\mid E_i^{gr} \mid,
  \label{eq:epsilon}
\ee

     with the requirement that all excess mass of the gas
component in star forming particle, which provide the
inequality $ \mid E_i^{gr} \mid > E_i^{th} + E_i^{ch} $, is
transformed into the star component. In the code we set the
absolute maximum value of the mass of such a "star"
particle $ m^{star}_{max} = 2.5 \cdot 10^6 \; M_\odot $,
\ie~ $ \approx 5 \% $ of the initial particle mass $ m_i $.

     At the moment of the birth, the position and velocities of new
"star" particles are equal to those of parent "gas" particles.
Thereafter these "star" particles interact with other "gas" and
"star" particles only by gravitation. The gravitational smoothing
length for these (Plummer like) particles is set equal to $
h_{star} $.


\subsection{The thermal SNII feed - back}

     We try to include in our code the events of SNII, SNIa
and PN into the complex gasdynamic picture of galaxy
evolution. But, for the thermal budget of the ISM, only
SNII play the main role. Following
Katz (1992), we assume that the explosion energy is
converted totally to the thermal energy. The stellar wind
action seems not to be essential in the energy budget
\cite{F95}. The total energy released by SNII explosions ($
10^{44} $ J per SNII) within a "star" particles is
calculated at each time step and distributed uniformly
between the surrounding (\ie~ $ r_{ij} < h_{star} $) "gas"
particles \cite{RVN96}.


\subsection{The chemical enrichment of gas}

     Every new "star" particle in our SF scheme represents
a separate, gravitationally closed star formation macro
region (like a globular cluster). The "star" particle is
characterized its own time of birth $ t_{begSF} $ which is
set equal to the moment of particle formation. After the
formation, these particles return the chemically enriched
gas to surrounding "gas" particles due to SNII, SNIa and PN
events. For the description of this process we use the
approximation proposed by
Raiteri \etal~ (1996). We consider only the production of $^{16}$O
and $^{56}$Fe, and try to describe the full galactic time
evolution of these elements, from the beginning to present time
(\ie~ $ t_{evol} \approx 13.0 $ Gyr).

    We use the multi - power IMF law defined by
Kroupa \etal~ (1993). The distribution of stellar masses within a
"star" particle of mass $ m^{star} $ is then:

\bea
   \Psi(m) = m^{star} \cdot A \cdot
            \left\{
            \begin{array}{lllll}
            2^{0.9} \cdot m^{-1.3}, & \mbox{ if $ 0.1  \leq m < 0.5 $}, \\
                                                                        \\
            m^{-2.2},               & \mbox{ if $ 0.5  \leq m < 1.0 $}, \\
                                                                        \\
            m^{-2.7},               & \mbox{ if $ 1.0  \leq m $}.       \\
            \end{array}
            \right.
   \label{eq:def_Psi}
\eea

     Here $ m $ is the star mass in solar units and the normalization
constant is $ A = 0.310146 $. The value of $ A $ is different from
Raiteri \etal~ (1996), because we set the integration limits $
m_{low} = 0.1 \; M_\odot $ and $ m_{upp} = 100 \; M_\odot $.

     For the definition of stellar lifetimes we use the equation
\cite{RVN96}:

\be
  \log t_{dead} = a_0({\rm Z}) - a_1({\rm Z}) \cdot \log m + a_2({\rm Z}) \cdot (\log m)^2,
  \label{eq:def_tdead}
\ee

     where $ t_{dead} $ is expressed in years and $ m $ is in solar unit
and coefficients are defined as:

\bea
  \left\{
  \begin{array}{lllll}
  a_0({\rm Z}) = 10.130 + 0.0755 \cdot \log {\rm Z} - 0.0081 \cdot (\log {\rm Z})^2, \\
                                  \\
  a_1({\rm Z}) = 4.4240 + 0.7939 \cdot \log {\rm Z} + 0.1187 \cdot (\log {\rm Z})^2, \\
                                  \\
  a_2({\rm Z}) = 1.2620 + 0.3385 \cdot \log {\rm Z} + 0.0542 \cdot (\log {\rm Z})^2. \\
  \end{array}
  \right.
  \label{eq:aZ}
\eea

     These relations are based on the calculations of the
Padova group \cite{ABB93,BFBC93,BBCFN94} and give a
reasonable approximation to stellar lifetimes in the mass
range from $ 0.6 \; M_\odot $ to $ 120 \; M_\odot $ and
metalicities Z $ = 7 \cdot 10^{-5} - 0.03 $ (defined as a
mass of all elements heavier than He). In our calculation,
we assume following,
Raiteri \etal~ (1996), that Z scales with the oxygen
abundance as Z/Z$_\odot = ^{16}$O$/^{16}$O$_\odot$. For
lower or higher metalicities we take the value
corresponding to the extremes of this interval.

    We can define the number of SNII explosions inside a given "star"
particle during the time from $ t $ to $ t + \Delta t $ using a
simple equation:

\be
  \Delta N_{{\rm SNII}} = \int\limits_{m_{dead}(t + \Delta t)}^{m_{dead}(t)} \Psi(m) dm.
  \label{eq:def_NSNII}
\ee

    Here $ m_{dead}(t) $ and $ m_{dead}(t + \Delta t) $ are masses of
stars that end their lifetimes at the beginning and at the end of
the time step respectively. Contrary to
Raiteri \etal~ (1996), we assume that all stars with masses
between $ 8 \; M_\odot $ and $ 100 \; M_\odot $ end their lives as
SNII. For SNII we use the yields from
Woosley \& Weaver (1995). We use the approximation formulae
from
Raiteri \etal~ (1996) for defining the total ejected mass by one
SNII - $ m^{tot}_{ej} $, as well as for the ejected mass of iron -
$ m^{{\rm Fe}}_{ej} $ and oxygen - $ m^{{\rm O}}_{ej} $ as a
function of stellar mass (in solar unit).

\bea
  \left\{
  \begin{array}{lllll}
  m^{tot}_{ej} = 7.682 \cdot 10^{-1} \cdot m^{1.056}, \\
  \\
  m^{{\rm Fe}}_{ej}  = 2.802 \cdot 10^{-4} \cdot m^{1.864}, \\
  \\
  m^{{\rm O}}_{ej}   = 4.586 \cdot 10^{-4} \cdot m^{2.721}. \\
  \end{array}
  \right.
  \label{eq:m_SNII}
\eea

     To take into account PN events inside the "star" particle we use the
equation, like (\ref{eq:def_NSNII}):

\be
  \Delta N_{{\rm PN}} = \int\limits_{m_{dead}(t + \Delta t)}^{m_{dead}(t)} \Psi(m) dm.
  \label{eq:def_NPN}
\ee

      Following
van den Hoek \& Groenewegen (1997), Samland (1997) and Samland
\etal~ (1997), we assume that all stars with masses between $ 1 \;
M_\odot $ and $ 8 \; M_\odot $ end their lives as PN. We define
the average ejected masses (in solar unit) of one PN event as
\cite{RV81,vdHG97}:

\bea
  \left\{
  \begin{array}{lllll}
  m^{tot}_{ej} = 1.63, \\
                       \\
  m^{{\rm Fe}}_{ej} = 0.00,  \\
                       \\
  m^{{\rm O}}_{ej} = 0.00.   \\
  \end{array}
  \right.
  \label{eq:m_PN}
\eea

     The method described in
Raiteri \etal~ (1996) and proposed in
Greggio \& Renzini (1983) and Matteuchi \& Greggio (1986) is used
to account of SNIa . In simulations, the number of SNIa exploding
inside a selected "star" particle during each time step is given
by:

\be
  \Delta N_{{\rm SNIa}} = \int\limits_{m_{dead}(t + \Delta t)}^{m_{dead}(t)} \Psi_2(m_2) dm_2.
  \label{eq:def_NSNIa}
\ee

     The quantity $ \Psi_2(m_2) $ represents the initial mass function
of the secondary component and includes the distribution function
of the secondary's mass relative to the total mass of the binary
system, $ m_{B} $,

\be
  \Psi_2(m_2) = m^{star} \cdot A_2 \cdot
  \int\limits_{m_{inf}}^{m_{sup}} (\frac{m_2}{m_B})^2 \cdot m_B^{-2.7} dm_B,
  \label{eq:def_Psi2}
\ee

     where $ m_{inf} = \max(2 \cdot m_2, \; 3 \; M_\odot) $ and $
m_{sup} = m_2 + 8 \; M_\odot $. The value of normalization
constant we set, following the
van den Berg \& McClure (1994), equal to $ A_2 = 0.16 \cdot A $.

     Then the total ejected mass (in solar unit) is \cite{TNY86,NTY84}:

\bea
  \left\{
  \begin{array}{lllll}
  m^{tot}_{ej} = 1.41, \\
                       \\
  m^{{\rm Fe}}_{ej} = 0.63,  \\
                       \\
  m^{{\rm O}}_{ej} = 0.13.   \\
  \end{array}
  \right.
  \label{eq:m_SNIa}
\eea

    In summary, a new "star" particle (with metalicity
Z $ = 10^{-4} $) with mass $ 10^4 \; M_\odot $, during the
total time of evolution $ t_{evol} $ produces:

$$ \Delta N_{{\rm SNII}} \approx 52.5, \; \Delta N_{{\rm PN}}
\approx 1770.0, \; \Delta N_{{\rm SNIa}} \approx 8.48. $$

\begin{figure}[htbp]

\vsp \insertplot{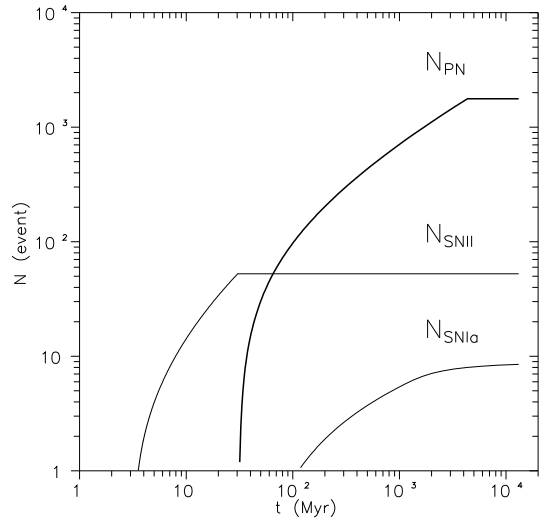}

\caption{The number of SNII, SNIa and PN event}

\label{fig-n}
\end{figure}

\begin{figure}[htbp]

\vsp \insertplot{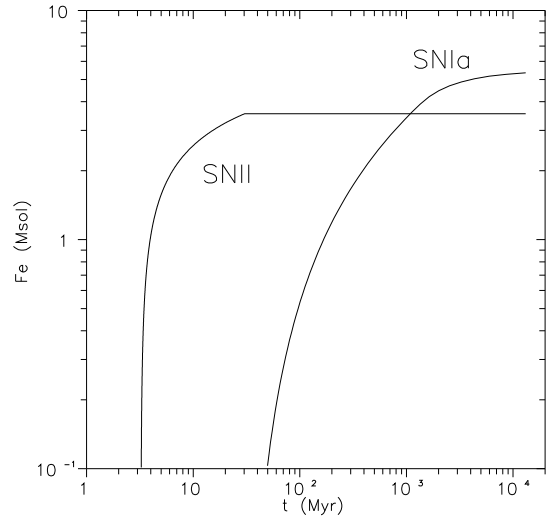}

\caption{The returned mass of $^{56}$Fe}

\label{fig-fe}
\end{figure}

\begin{figure}[htbp]

\vsp \insertplot{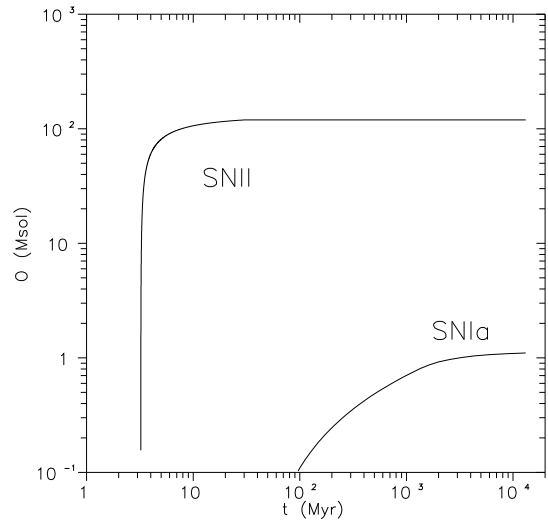}

\caption{The returned mass of $^{16}$O}

\label{fig-o}
\end{figure}

     We present in Fig.~\ref{fig-n}. the number of SNII, SNIa and PN
events for this "star" particle.

     In Fig.~\ref{fig-fe}. and Fig.~\ref{fig-o}. we present the returned
masses of $^{56}$Fe and $^{16}$O, respectively. We can estimates
the total masses (H, He, $^{56}$Fe, $^{16}$O) (in solar masses)
returned to the surrounding "gas" particles, due to these
processes as:

\bea
  \left\{
  \begin{array}{lllllll}
  \Delta m^{{\rm H}}_{{\rm SNII}}  = 477, \;\;\; \Delta m^{{\rm H}}_{{\rm PN}}  = 2164,\;\;\;\; \Delta m^{{\rm H}}_{{\rm SNIa}}  = 4.14, \\
  \\
  \Delta m^{{\rm He}}_{{\rm SNII}} = 159, \;\;\; \Delta m^{{\rm He}}_{{\rm PN}} = 721.3, \;\;\; \Delta m^{{\rm He}}_{{\rm SNIa}} = 1.38, \\
  \\
  \Delta m^{{\rm Fe}}_{{\rm SNII}} = 3.5, \;\;\; \Delta m^{{\rm Fe}}_{{\rm PN}} = 0.000, \;\;\; \Delta m^{{\rm Fe}}_{{\rm SNIa}} = 5.35, \\
  \\
  \Delta m^{{\rm O}}_{{\rm SNII}}  = 119, \;\;\; \Delta m^{{\rm O}}_{{\rm PN}}  = 0.000, \;\;\; \Delta m^{{\rm O}}_{{\rm SNIa}}  = 1.10. \\
  \end{array}
  \right.
  \label{eq:masses}
\eea


\subsection{The cold dark matter halo}

     In the literature we have found some, sometimes controversial,
profiles for the galaxies Cold Dark Matter Haloes (CDMH)
\cite{B95,N98}. For resolved structures of CDMH: $
\rho_{halo}(r) \sim r^{-1.4} $ \cite{MGQSL97}. The
structure of CDMH high-resolution N-body simulations show
can be described by: $ \rho_{halo}(r) \sim r^{-1} $
\cite{NFW96,NFW97}. Finally, in
Kravtsov \etal~ (1997), we find that the cores of DM dominated
galaxies may have a central profile: $ \rho_{halo}(r) \sim
r^{-0.2} $.

     Because we concentrate our attention mainly on the
problem of the formation and evolution (dynamical and
chemical) of the disk structure in galaxy we using the
simplest dark matter halo model. In our calculations, as a
first approximation, it is assumed that the model galaxy
halo contains the CDMH component with Plummer - type
density profiles \cite{DC95}:

\be
\rho_{halo}(r) = \frac{ M_{halo} }{ \frac{4}{3} \pi b_{halo}^3 }
\cdot
                 \frac{ b_{halo}^5 }{ (r^2 + b_{halo}^2 )^{\frac{5}{2}} },
\ee

     Therefore we can write for the external force acting on the "gas"
and "star" particles:

\be
- \nabla_i \Phi^{ext}_i = - G \cdot \frac{ M_{halo} }
                            { ({\bf r}^2_{i} + b^2_{halo})^{\frac{3}{2}} }
                            \cdot {\bf r}_{i}.
\ee

     The inclusion of the dynamically evolved halo model is a
next step in our calculations.


\section{Results and discussion}

\subsection{Initial conditions}

     We test our code and demonstrate that simple assumptions lead to a
reasonable galaxy model. The SPH calculations were carried out for
$ N_{gas} = 2109 $ "gas" particles. According to
Navarro \& White (1993) and Raiteri \etal~ (1996), such
number seems to be quite enough to provide a qualitatively
correct description of the system behaviour. Even such
small number of "gas" particles produces $ N_{star} = 31631
$ "star" particles at the end of the calculation. We also
check the dependency of our model with the number of "gas"
particles. We calculate the first $\sim$ 5 Gyr evolution of
the systems with identical initial data but with different
numbers of $N_{gas} = 2109; \; 4169; \; 11513$ with
correspondent $N_B = 21; \; 41; \; 115$. All three models
show the practically identical star formation rate and
chemical evolution history.

     The value of the smoothing length $ h_i $ was chosen
requiring that each "gas" particle had $ N_B = 21 $
neighbors within $ 2 \cdot h_i $. Minimal $ h_{min} $ was
set equal to $ 1 $ kpc. This value is defined as a largest
characteristic dimensions of instability in galactic disk.
This $h_{min}$ is enough only for a crude description of
the galaxy disk structure but is quite enough for a
dimensions of SSP (Single Stellar Population) in chemical
evolution modeling. Any dynamical processes below this
scale we can't describe correctly in our SPH code. We use
also the fixed gravitational smoothing length $ h_{star} =
1 $ kpc for the "star" particles. Our results show that
such a value of $ N_B \approx 1 \% \; N_{gas} $ provides
qualitatively correct treatment of the system's large scale
evolution.

     As the initial model (relevant for CDM - scenario)
we took constant - density homogeneous gaseous triaxial
configuration ($ M_{gas} = 10^{11} \; M_\odot $) within the
dark matter halo ($ M_{halo} = 10^{12} \; M_\odot $). We
set $ A = 100 $ kpc, $ B = 75 $ kpc and $ C = 50 $ kpc for
semiaxes of system. Such a triaxial configurations are
reported in cosmological simulations of the dark matter
halo formation \cite{EL95,FWDE88,WQSZ92}. Initially, the
centers of all particles were placed on a homogeneous grid
inside this triaxial configuration. We set the smoothing
parameter of CDMH: $ b_{halo} = 25 $ kpc. Such values of $
M_{halo} $ and $ b_{halo} $ are typical for CDMH in disk
galaxies \cite{NFW96,NFW97,B95}.

     The gas component was assumed to be initially cold,
$ T_0 = 10^{4} $ K. As we see in our calculations, the
influence of chaotic motions essentially reduces the
dependence of model parameters on the adopted temperature
cutoff and, therefore, on the adopted form of cooling
function itself.

     The gas was assumed to be involved into the Hubble
flow ($ H_{0} = 65 $ km/s/Mpc) and into the solid - body
rotation around $ z $ - axis. We added the small random
components of velocities ($ \Delta \mid {\bf v} \mid = 10 $
km/s) to account for the chaotic motions of fragments. The
initial velocity field was defined as:

\be
   {\bf v}(x, y, z) = [{\bf \Omega}(x, y, z) \times {\bf r}]
                         + H_{0} \cdot {\bf r}
                         + \Delta {\bf v}(x, y, z),
   \label{eq:V_0}
\ee

     where $ {\bf \Omega}(x,y, z) $ is the angular velocity
of an initially rigidly rotating system.

     The spin parameter in our numerical simulations
is $ \lambda \approx 0.08 $. This parameter is defined in
Peebles (1969) as:

\begin{equation}
  \lambda = \frac{\mid {\bf L}_0 \mid \cdot \sqrt{\mid E_0^{gr} \mid}}
              {G \cdot (M_{gas}+M_{halo})^{5/2}},
   \label{eq:Spin}
\end{equation}

     $ {\bf L}_0 $ is the total initial angular momentum
and $ E_0^{gr} $ is the total initial gravitational energy
of a protogalaxy. It is to be noted that for a system, in
which angular momentum is acquired through the tidal torque
of the surrounding matter, the standard spin parameter does
not exceed $ \lambda \approx 0.11 $ \cite{SB95}. Moreover,
its typical values range between $ \lambda \approx
0.07^{+0.04}_{-0.05} $, \eg~ $ 0.02 \le \lambda \le 0.11 $.


\subsection{Dynamical model}

     At the final time step ($ t_{evol} \approx 13.0 $ Gyr)
the "star" and "gas" particle distributions have dimensions
typical for a disk galaxy. The radial extension is
approximately $ 25 - 30 $ kpc. The disk height is around $
1 - 2 $ kpc. The "gas" particles are located mainly within
the central $ 5 - 10 $ kpc.

\begin{figure}[htbp]

\vsp \insertplot{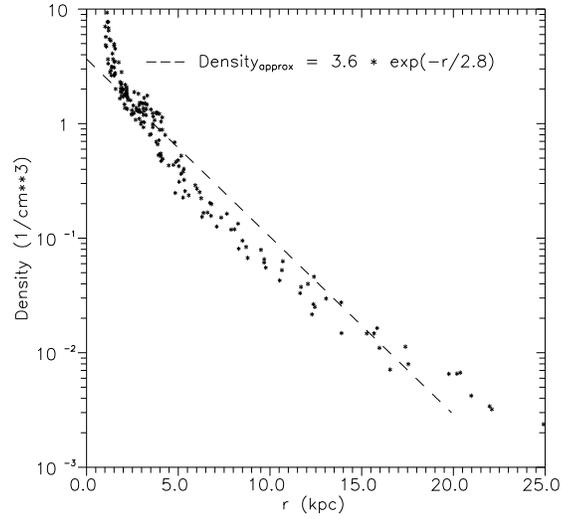}

\caption{$ \rho(r) $. The density distribution of gas in the final
step}

\label{fig-den}
\end{figure}

     We present in Fig.~\ref{fig-den}. the density distribution
of gas $ \rho(r) $. Except for the central region ($ < 2 $
kpc), the gas distribution has a exponential form with
radial scale length $ \approx 2.8 $ kpc.

\begin{figure}[htbp]

\vsp \insertplot{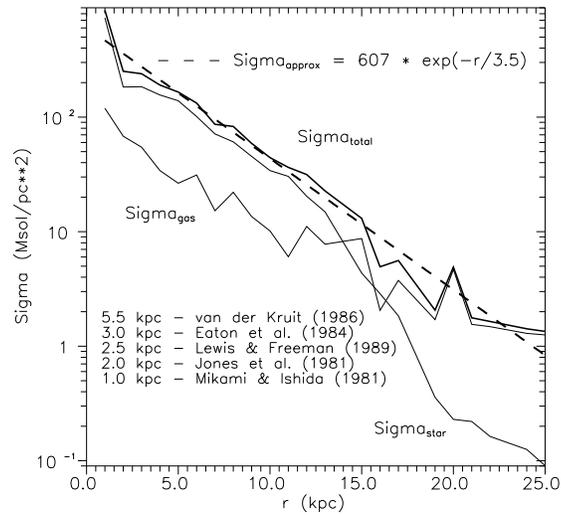}

\caption{$ \sigma_{gas}(r) $,~~~$ \sigma_{*}(r) $ and $
\sigma_{tot}(r) = \sigma_{gas}(r) + \sigma_{*}(r) $. The column
density distribution in the disk of gas and stars in the final
step}

\label{fig-sigma}
\end{figure}

     The column density distributions of gas $ \sigma_{gas}(r) $
and stars $ \sigma_{*}(r) $ are presented in
Fig.~\ref{fig-sigma}. The total column density is defined
as: $ \sigma_{tot}(r) = \sigma_{gas}(r) + \sigma_{*}(r) $.
The total column density distribution $ \sigma_{tot}(r) $
is well approximated (in the interval from $ 5 $ kpc to $
15 $ kpc) with an exponential profile characterised with a
$ \approx 3.5 $ kpc radial scale length.

     In the literature we found a large scatter of values for
this radial scale length obtained for our Galaxy:

\begin{itemize}
\item $ 5.5 $ kpc \cite{vdK86},
\item $ 3.0 $ kpc \cite{EAG84},
\item $ 2.5 $ kpc \cite{LF89},
\item $ 2.0 $ kpc \cite{JAHR81},
\item $ 1.0 $ kpc \cite{MI81},
\item $ 2.1 \pm 0.3 $ kpc \cite{PGJB98},
\item $ 2.3 \pm 0.1 $ kpc \cite{RRECBFG96},
\item $ 2.5^{+0.8}_{-0.6} $ kpc \cite{FM94}.
\end{itemize}

     Our value is very close with a recently reported
value: $ 3.5 $ kpc \cite{MCS98b}.

     The value of $ \sigma_{tot} \approx 55 \; M_\odot/pc^2 $
near the location of the Sun ($ r \approx 9 $ kpc) is very
close to a recent determination of the total density $ 52
\pm 13 \; M_\odot/pc^2 $ \cite{MCS98a}.

\begin{figure}[htbp]

\vsp \insertplot{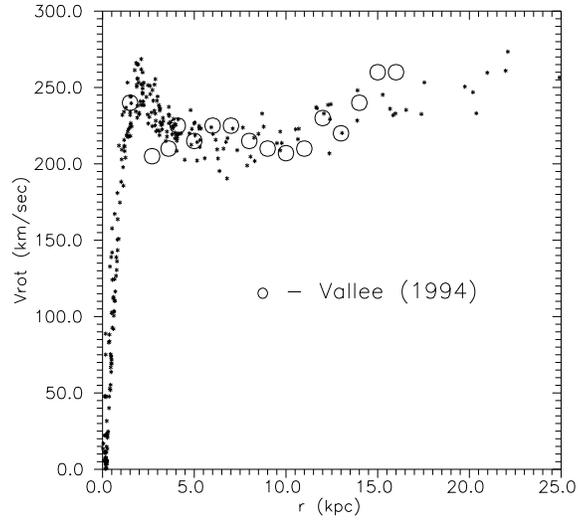}

\caption{$ V_{rot}(r) $. The rotational velocity distribution of
gas in the final step}

\label{fig-v_rot}
\end{figure}

     We present in Fig.~\ref{fig-v_rot}. the rotational velocity
distribution of gas $ V_{rot}(r) $, at the end of calculation. In
the figure we show the rotational curve for our Galaxy
\cite{Va94}. As we can see from the figure, the modelled disk
galaxy rotation velocities are very close to our Galaxy rotation
curve.

\begin{figure}[htbp]

\vsp \insertplot{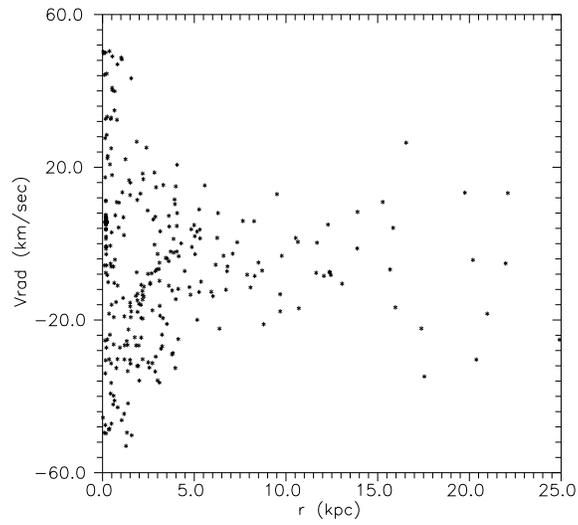}

\caption{$ V_{rad}(r) $. The radial velocity distribution of gas
in the final step}

\label{fig-v_rad}
\end{figure}

\begin{figure}[htbp]

\vsp \insertplot{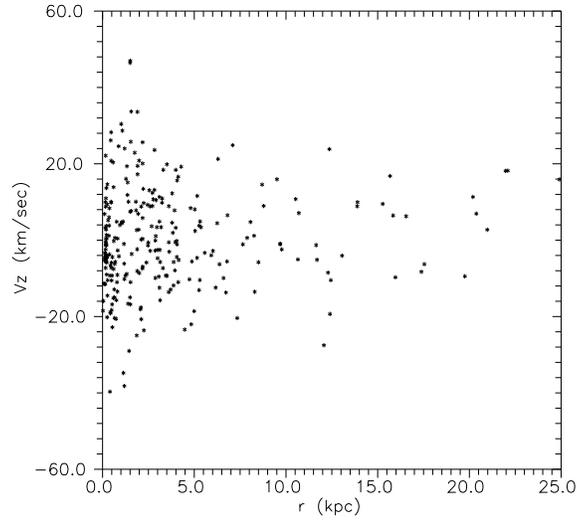}

\caption{$ V_{z}(r) $. The perpendicular to disk normal velocity
distribution of gas in the final step}

\label{fig-v_z}
\end{figure}

     The gaseous radial $ V_{rad}(r) $ and normal $ V_{z}(r) $
velocity distributions we present in Fig.~\ref{fig-v_rad}.
and Fig.~\ref{fig-v_z}. The radial velocity dispersion has
a maximum value $ \approx 60 $ km/s in the center (this
high value is mainly caused by the central strong bar
structure). Near the Sun this dispersion drops down to $
\approx 20 $ km/s. Such radial dispersion is reported in
the kinematics study of the stellar motions in the solar
neighborhood \cite{B98}. The normal dispersion is near $
\approx 20 $ km/s in the whole disk. This value also
coincides with the vertical dispersion velocity near the
Sun \cite{B98}.

\begin{figure}[htbp]

\vsp \insertplot{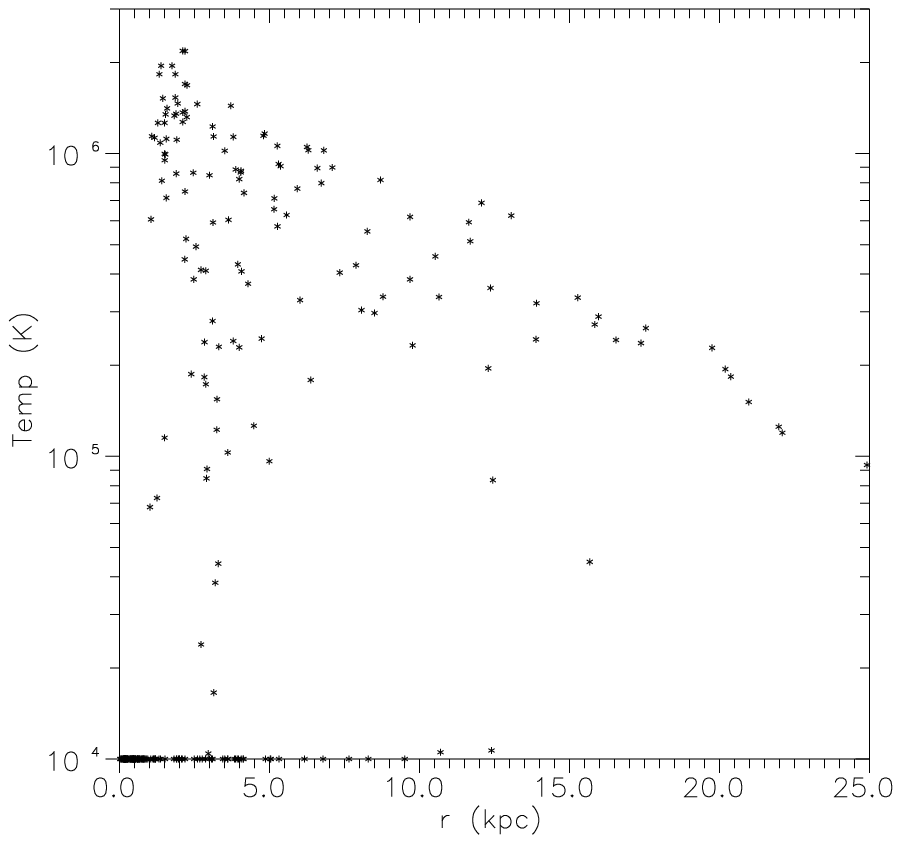}

\caption{$ T(r) $. The temperature distribution of gas in
the final step}

\label{fig-temp}
\end{figure}

     We present in Fig.~\ref{fig-temp}. the temperature
distribution of gas $ T(r) $. As we can see from the
figure, the distribution of $ T(r) $ has a very large
scatter from $ 10^4 $ K to $ 10^6 $ K. Because, in our
calculation we set the cutoff temperature for the cooling
function at $ 10^4 $ K, the gas can't cool to lower
temperatures. The most ($\sim$ 90 \%) of the gaseous mass
in the galaxy have a temperatures $ \sim 10^4 $ K. The
modeled process of SNII explosions injects to the gas a
great amount of thermal energy and generates a very large
temperature scatter. Such a temperature scatter is typical
for our Galaxy's ISM.

     Therefore, as we can see, even our crude numerical
approximation gives a good fit for all dynamical and
thermal distributions of gas and stars in a typical disk
galaxy like our Galaxy.

\subsection{Chemical characteristics}

\begin{figure}[htbp]

\vsp \insertplot{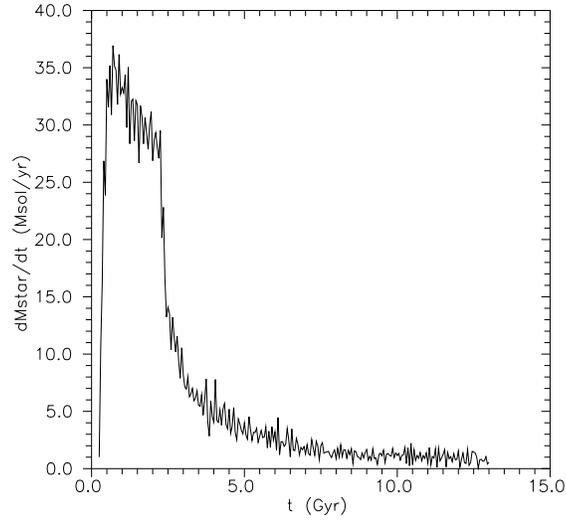}

\caption{$ SFR(t) = dM_{*}(t)/dt $. The time evolution of the SFR
in galaxy}

\label{fig-dm}
\end{figure}

     We present in Fig.~\ref{fig-dm}. the time evolution
of the SFR in galaxy $ SFR(t) = dM_{*}(t)/dt $. About $
\approx 90 \% $ of gas is converted into stars at the end
of calculation. The most intensive SF burst happened in the
first $ \approx 1 $ Gyr, with a maximal SFR $ \approx 35 \;
M_\odot/yr $. After $ \approx 1.5 - 2 $ Gyr the SFR is
coming down like an "exponential function". The SFR has a
value $ \approx 1 \; M_\odot/yr $ at the end of the
simulation.

     To check the SF and chemical enrichment algorithm in our
SPH code, we use the chemical characteristics of the disk
in the "solar" cylinder ($ 8 $ kpc $ < \; r \; < \; 10 $
kpc).

\begin{figure}[htbp]

\vsp \insertplot{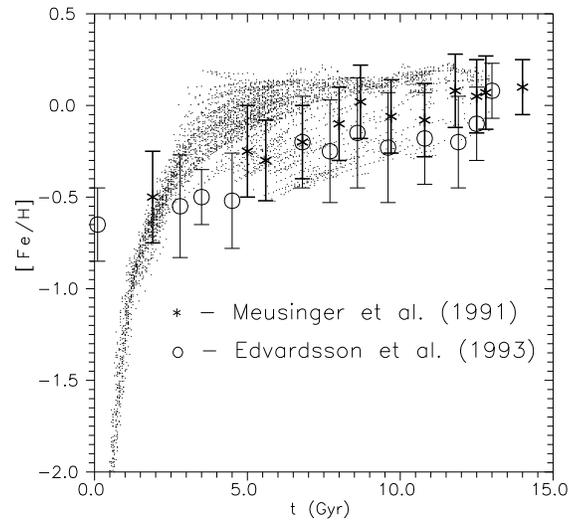}

\caption{[Fe/H]$(t)$. The age metalicity relation of the "star"
particles in the "solar" cylinder ($ 8 $ kpc $ < \; r \; < \; 10 $
kpc)}

\label{fig-chem_1}
\end{figure}

     The age - metalicity relation of the "star" particles in the
"solar" cylinder, [Fe/H]$(t)$, we present in
Fig.~\ref{fig-chem_1}. The observational data in this figure are
taken from
Meusinger \etal~ (1991) and Edvardsson \etal~ (1993).

\begin{figure}[htbp]

\vsp \insertplot{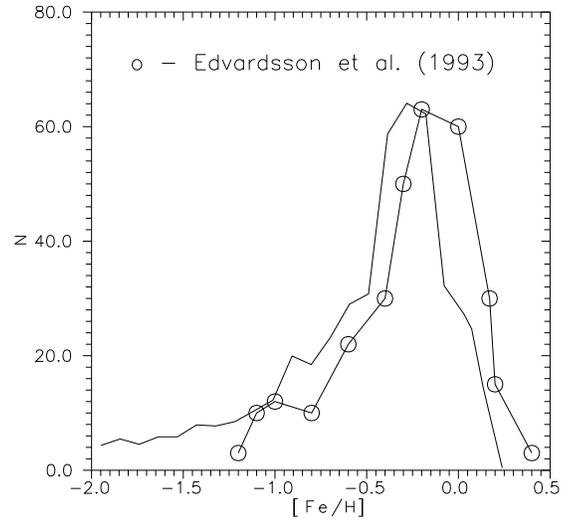}

\caption{$N_{*}($[Fe/H]$)$. The metalicity distribution of
the "star" particles in the "solar" cylinder ($ 8 $ kpc $ <
\; r \; < \; 10 $ kpc)}

\label{fig-chem_2}
\end{figure}

     We presented in Fig.~\ref{fig-chem_2}. the metalicity
distribution of the "star" particles in the "solar"
cylinder $N_{*}($[Fe/H]$)$. The model data are scaled to
the observed number of stars \cite{EAG93}.

\begin{figure}[htbp]

\vsp \insertplot{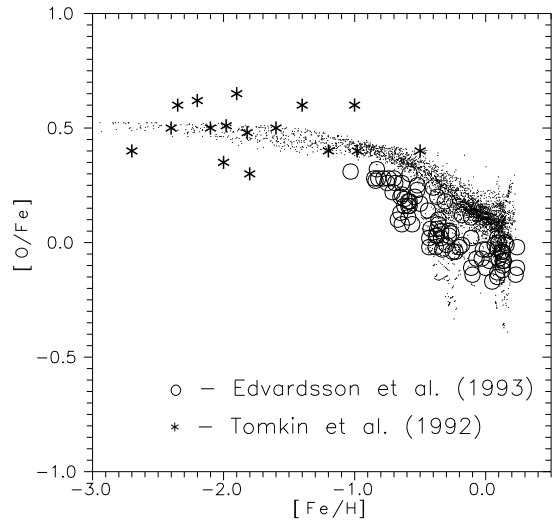}

\caption{The [O/Fe] vs. [Fe/H] distribution of the "star"
particles in the "solar" cylinder ($ 8 $ kpc $ < \; r \; < \; 10 $
kpc)}

\label{fig-chem_3}
\end{figure}

     The [O/Fe] vs. [Fe/H] distribution of the "star" particles in the
"solar" cylinder we presented in Fig.~\ref{fig-chem_3}. In this
figure we also present the observational data from
Edvardsson \etal~ (1993) and Tomkin \etal~ (1992).

     All these model distributions are in good agreement, not only with
presented observational data, but also with other data collected
from
Portinari \etal~ (1997).

\begin{figure}[htbp]

\vsp \insertplot{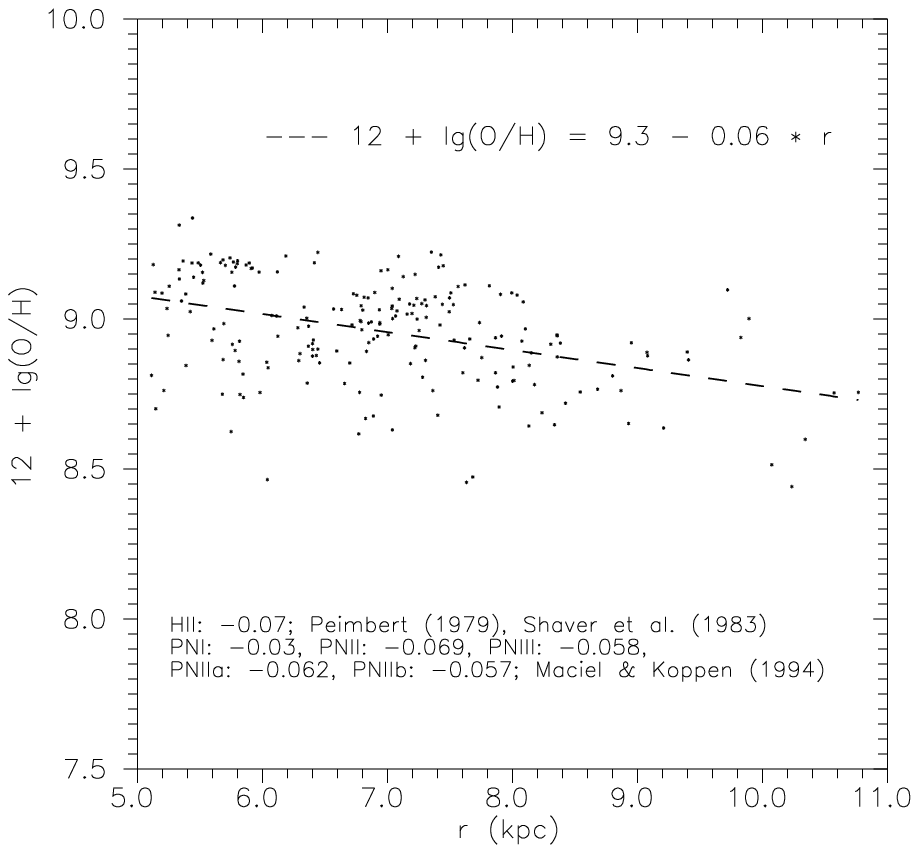}

\caption{[O/H]$(r)$. The [O/H] radial distribution}

\label{fig-grad-z}
\end{figure}

     The [O/H] radial distribution [O/H]$(r)$ we present in
Fig.~\ref{fig-grad-z}. The approximation presented in the
figure is obtained by a least - squares linear fit. At
distances $ 5 $ kpc $ < \; r \; < \; 11 $ kpc the model
radial abundance gradient is $ -0.06 $ dex/kpc. In the
literature we found different values of this gradient
defined in objects of different types. From observations of
HII regions \cite{P79,SMNDP83} we obtained that the oxygen
radial gradient is $ -0.07 $ dex/kpc. From observations of
PN of different types \cite{MK94} we obtained the values: $
-0.03 $ dex/kpc for PNI, $ -0.069 $ dex/kpc for PNII, $
-0.058 $ dex/kpc for PNIII, $ -0.062 $ dex/kpc for PNIIa, $
-0.057 $ dex/kpc for PNIIb. Therefore our model agrees well
with the oxygen radial gradient in the our Galaxy.

\subsection{Conclusion}

     This simple model provides a reasonable, self - consistent
picture of the process of galaxy formation and star
formation in the galaxy. Our calculations show that even a
relatively small number of "gas" particles with physically
motivated star formation criteria can reproduce the most of
observed dynamical and chemical characteristic in the
galactic disks. The dynamical and chemical evolution of the
modeled disk - like galaxy is coincident with the results
of observations for our own Galaxy. The results of our
modeling give a good base for a wide use of the proposed SF
and chemical enrichment algorithm in other SPH simulations.


\bigskip

     {\bf Acknowledgments.} The author is grateful to S.G. Kravchuk,
L.S. Pilyugin and Yu.I. Izotov for fruitful discussions during the
preparation of this work. It is pleasure to thank Pavel Kroupa and
Christian Boily for comments on an earlier version of this paper.
The author also thank the anonymous referee for a helpful
referee's report which resulted in a significantly improved
version of this paper. This research was supported by a grant from
the American Astronomical Society.


\newpage




\end{document}